\documentstyle[epsfig]{aipproc}

\def\MPL #1 #2 #3 {Mod.~Phys.~Lett.~{\bf#1},\  #2 (#3)}
\def\NPB #1 #2 #3 {Nucl.~Phys.~{\bf #1},\  #2 (#3)}
\def\PLB #1 #2 #3 {Phys.~Lett.~{\bf #1},\  #2 (#3)}
\def\PR #1 #2 #3 {Phys.~Rep.~{\bf#1},\ #2 (#3)}
\def\PRD #1 #2 #3 {Phys.~Rev.~{\bf #1},\  #2 (#3)}
\def\PRL #1 #2 #3 {Phys.~Rev.~Lett.~{\bf#1},\  #2 (#3)}
\def\RMP #1 #2 #3 {Rev.~Mod.~Phys.~{\bf#1},\  #2 (#3)}
\def\ZPC #1 #2 #3 {Z.~Phys.~{\bf #1},\  #2 (#3)}
\def\IJMP #1 #2 #3 {Int.~J.~Mod.~Phys.~{\bf#1},\  #2 (#3)}

\def\fnal97{FNAL$\mu\mu$97}

\def\bit{\begin{itemize}}
\def\eit{\end{itemize}}
\def\pzero{P^0}
\def\mpzero{m_{\pzero}}

\newcommand{\beq}{\begin{equation}}
\newcommand{\eeq}{\end{equation}}
\newcommand{\bea}{\begin{eqnarray}}
\newcommand{\eea}{\end{eqnarray}}
\newcommand{\nn}{\nonumber}

\def\Eq#1{Eq.~(\ref{#1})}

\def\lsim{\mathrel{\raise.3ex\hbox{$<$\kern-.75em\lower1ex\hbox{$\sim$}}}}
\def\gsim{\mathrel{\raise.3ex\hbox{$>$\kern-.75em\lower1ex\hbox{$\sim$}}}}

\def\to{\rightarrow}

\def\mupmum{\mu^+\mu^-}

\def\gev{~{\rm GeV}}

\def\rp{\not{\hbox{\kern-4pt $R_P$}}}
\def\met{\not{\hbox{\kern-4pt $E_T$}}}
\def\mpt{\not{\hbox{\kern-4pt $p_T$}}}
\def\fbi{ \text{ fb}^{-1}}
\def\anti{\overline}
\def\rts{\sqrt s}
\def\sigrts{\sigma_{\!\rts}}

\def\To{\Rightarrow}

\def\etc{{\it etc.}}

\def\etc{{\em etc.}}

\def\cale{{\cal E}}
\def\calo{{\cal O}}
\def\eg{{\it e.g.}}

\def\mstop{m_{\stop}}

\def\stop{\wt t}

\def\hsm{h_{\rm SM}}
\def\mhsm{m_{\hsm}}
\def\hl{h^0}
\def\hh{H^0}
\def\ha{A^0}
\def\hp{H^+}
\def\hm{H^-}
\def\hpm{H^{\pm}}
\def\mhl{m_{\hl}}
\def\mhh{m_{\hh}}
\def\mha{m_{\ha}}

\def\mhpm{m_{\hpm}}
\def\tanb{\tan\beta}

\def\mz{m_Z}
\def\mw{m_W}

\def\wm{W^-}

\def\h{h}

\def\mh{m_{\h}}

\def\dmm{\Delta^{--}}
\def\mdmm{m_{\dmm}}

\def\gamdmm{\Gamma_{\dmm}^{\rm tot}}

\def\mVV{M_{VV}}

\def\em{e^-}

\def\hpm{H^{\pm}}

\def\call{{\cal L}}
\def\cmsi{{\rm cm}^{-2}{\rm s}^{-1}}

\def\tauptaum{\tau^+\tau^-}
\def\mbb{m_{b\anti b}}

\def\ltot{L_{\rm tot}}

\def\lam{\lambda}

\def\br{B}
\def\tauptaum{\tau^+\tau^-}
\def\mbb{m_{b\anti b}}

\def\shat{{\hat s}}
\def\rtshat{\sqrt{\shat}}
\def\gam{\gamma}
\def\sigrts{\sigma_{\tiny\rts}^{}}

\def\etc{{\it etc.}}
\def\sighbar{\overline \sigma_{\h}}

\def\sighsmbar{\overline\sigma_{\hsm}}

\def\anti{\overline}
\def\epem{e^+e^-}
\def\zstar{Z^\star}
\def\wstar{W^\star}

\def\mupmum{\mu^+\mu^-}

\def\rts{\sqrt s}
\def\ie{{\it i.e.}}
\def\eg{{\it e.g.}}

\def\anti{\overline}

\def\wm{W^-}
\def\mw{m_W}
\def\mz{m_Z}
\def\h{h}
\def\mh{m_{\h}}
\def\gamh{\Gamma_{\h}^{\rm tot}}

\def\hsm{h_{SM}}
\def\mhsm{m_{\hsm}}
\def\gamhsm{\Gamma_{\hsm}^{\rm tot}}
\def\tanb{\tan\beta}
\def\hl{h^0}
\def\mhl{m_{\hl}}

\def\ha{A^0}
\def\mha{m_{\ha}}
\def\gamha{\Gamma_{\ha}^{\rm tot}}
\def\hh{H^0}
\def\mhh{m_{\hh}}
\def\gamhh{\Gamma_{\hh}^{\rm tot}}
\def\fbi{~{\rm fb}^{-1}}

\def\mev{~{\rm MeV}}
\def\gev{~{\rm GeV}}
\def\tev{~{\rm TeV}}
\def\stop{\widetilde t}
\def\mstop{m_{\stop}}

\def\gampzero{\Gamma_{\pzero}^{\rm tot}}

\def\tanb{\tan\beta}

\begin{document}
\font\fortssbx=cmssbx10 scaled \magstep2
\hbox to \hsize{
$\vcenter{
\hbox{\fortssbx University of California - Davis}
}$

\hfill
$\vcenter{
\hbox{\bf UCD-98-7} 
\hbox{\bf hep-ph/9804358}
\hbox{April, 1998}
}$
}
\medskip

\title{Higgs and Technicolor Goldstone Bosons at a Muon Collider
\thanks{To appear in {\it Proceedings of the 4th International
Conference on ``Physics Potential and Development of $\mupmum$
Colliders''}, 
San Francisco, December 8--10, 1997, eds. D. Cline and K. Lee.
This work was supported in 
part by U.S. Department of Energy grant No. DE-FG03-91ER40674.}
}
\author{John F. Gunion}
\address{Davis Institute for High Energy Physics,
Department of Physics, University of California,
Davis, CA 95616, USA}

\maketitle

\newlength{\captsize} \let\captsize=\small 

\def\mVV{M_{VV}}
\def\muc{$\mu$C}
\def\ec{$e$C}
\def\ellc{$\ell$C}

\begin{abstract}
I discuss the exciting prospects for Higgs and technicolor Goldstone boson 
physics  at a muon collider. 
\end{abstract}

\section*{Introduction}

The prospects for Higgs and Goldstone boson 
physics at a muon collider depend crucially upon
the instantaneous luminosity, $\call$, possible
for $\mupmum$ collisions as a function of $E_{\rm beam}$ and on the percentage
Gaussian spread in the beam energy, denoted by $R$.
The small level of bremsstrahlung and absence of beamstrahlung 
at muon collider implies
that very small $R$ can be achieved. The (conservative)
luminosity assumptions for the recent Fermilab-97 workshop were:~\footnote{For
yearly integrated luminosities, we use the standard convention
of $\call=10^{32}$cm$^{-2}$s$^{-1}\To L=1\fbi/{\rm yr}$.}
\begin{itemize}
\item
$\call\sim (0.5,1,6) \cdot 10^{31}\cmsi$ for $R=(0.003,0.01,0.1)\%$
at $\rts\sim 100\gev$;
\item
$\call\sim (1,3,7) \cdot 10^{32}\cmsi$,
at $\rts\sim (200,350,400)\gev$, $R\sim 0.1\%$.
\end{itemize}
With modest success in the collider design, at least a factor of 2
better can be anticipated. Note that
for $R\sim 0.003\%$ the Gaussian spread in $\rts$, given by
$\sigrts\sim 2\mev\left({R\over 0.003\%}\right)\left({\rts\over
100\gev}\right)$,  can be comparable to
the few MeV widths of very narrow resonances
such as a light SM-like Higgs boson or a (pseudo-Nambu-Goldstone)
technicolor boson.
This is critical since the effective resonance cross section $\overline\sigma$
is obtained by convoluting a Gaussian $\rts$ distribution of width
$\sigrts$ with the standard $s$-channel Breit Wigner resonance 
cross section, $\sigma(\rtshat)=
4\pi\Gamma(\mu\mu)\Gamma(X)/([\shat-M^2]^2+[M\Gamma^{\rm tot}]^2)$.
For $\rts=M$, the result,~\footnote{In actual
numerical calculations, bremsstrahlung smearing is
also included (see Ref.~\cite{bbgh}).}              
\begin{equation}
\overline\sigma\simeq {\pi\sqrt{2\pi} \Gamma(\mu\mu)\, \br( X) \over
M^2\sigrts}
\times \left(1+ {\pi\over 8}\left[{\Gamma^{\rm tot}\over
\sigrts}\right]^2\right)^{-1/2}\,, 
\label{sigmaform}
\end{equation}
will be maximal if $\Gamma^{\rm tot}$ is small and 
$\sigrts\sim\Gamma^{\rm tot}$.~\footnote{Although smaller 
$\sigrts$ (\ie\ smaller $R$) implies smaller $\call$, 
the $\call$'s given earlier are such that when $\Gamma^{\rm tot}$ 
is in the MeV range
it is best to use the smallest $R$ that can be achieved.}
Also critical to scanning a narrow resonance is the ability~\cite{raja}
to tune the beam energy to one part in $10^{6}$.

\section*{\bf Higgs Physics}

The potential of the muon collider for Higgs physics is truly outstanding.
First, it should be emphasized that 
away from the $s$-channel Higgs pole, $\mupmum$ and $\epem$ colliders
have similar capabilities for the same $\rts$ and $\call$ (barring
unexpected detector backgrounds at the muon collider). At $\rts=500\gev$,
the design goal for a $\epem$ linear collider (\ec) is $L=50\fbi$ per year.
The conservative $\call$
estimates given earlier suggest that at $\rts=500\gev$
the \muc\ will accumulate {\it at least} $L=10\fbi$ per year.  If this can
be improved somewhat, the \muc\ would be fully competitive with the \ec\
in high energy ($\rts\sim 500\gev$) running.
(We will use the notation of \ellc\ for either a \ec\ or \muc\ operating
at moderate to high $\rts$.)

The totally unique feature of the \muc\ is the dramatic peak in
the cross section $\sighbar$ for production of a narrow-width 
Higgs boson in the $s$-channel
that occurs when $\rts=\mh$ and $R$ is small enough that $\sigrts$
is smaller than or comparable to $\gamh$ \cite{bbgh}.
The peaking is illustrated below in Fig.~\ref{gausssigma}
for a SM Higgs ($\hsm$) with $\mhsm=110\gev$ ($\gamhsm\sim 3\mev$).

\medskip
\noindent{\bf A Standard Model-Like Higgs Boson}
\medskip

For SM-like $\h\to WW,ZZ$ couplings, $\gamh$ becomes big
if $\mh\gsim 2\mw$, and 
$\sighbar\propto \br(\h\to\mupmum)$  [\Eq{sigmaform}] will
be small; $s$-channel production will not be useful.
But, as shown in Fig.~\ref{gausssigma},
$\sighbar$ is enormous for small $R$ when the $\h$ is light, as is
very relevant in supersymmetric models where the light SM-like $\hl$ has
$\mhl\lsim 150\gev$.
In order to make use of this large cross section, we must
first center on $\rts\sim \mh$ and then proceed to
the precision measurement of the Higgs boson's properties.

\begin{figure}[h]
\begin{center}
{\epsfig{file=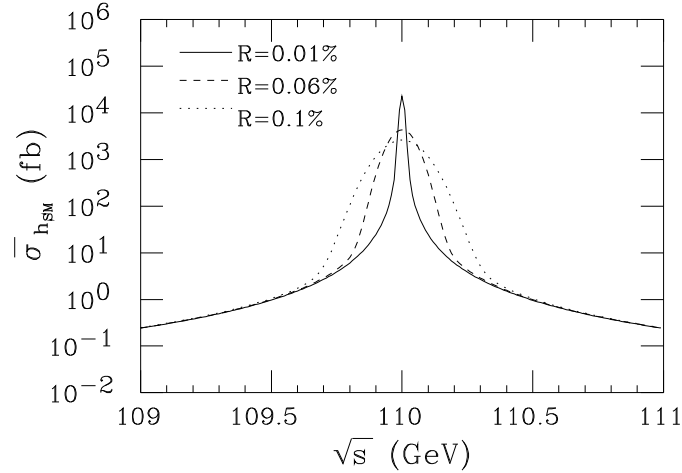,width=3.5in}}
\caption{ The effective cross section, $\sighsmbar$,
for $R=0.01\%$, $R=0.06\%$, and $R=0.1\%$ vs. $\protect\rts$ for 
$\mhsm=110\gev$.
\label{gausssigma}}
\end{center}
\end{figure}

For a SM-like Higgs with $\mh\lsim 2\mw$ one expects~\cite{jfghiggs}
to determine $\mh$ to within
$\Delta\mh\sim 100\mev$ from LHC data ($L=300\fbi$) (the uncertainty 
$\Delta\mh$ will be even smaller if \ellc\
data is available). Thus, a final ring that
is fully optimized for $\rts\sim\mh$ can be built.
Once it is operating, we scan over the appropriate $\Delta\mh$ interval so as
to center on $\rts\simeq\mh$ within a fraction of $\sigrts$.
Consider first the ``typical'' case of $\mh\sim 110\gev$.
For $\mh$ of order 100 GeV, $R=0.003\%$ implies $\sigrts\sim 2\mev$.
$\Delta\mh\sim 100\mev$ implies that $\Delta\mh/\sigrts \sim 50$ 
points are needed to center within $\lsim \sigrts$.  At this mass,
each point requires $L\sim 0.0015\fbi$ in order to observe
or eliminate the $\h$ at the $3\sigma$ level, implying a total of 
$\ltot\leq 0.075\fbi$ is needed for centering. (Plots as a function of $\mhsm$ 
of the luminosity required for a $5\sigma$ observation of 
the SM Higgs boson when $\rts=\mhsm$ can be found in Ref.~\cite{bbgh}.)
Thus, for the anticipated $L\sim 0.05-0.1\fbi/{\rm yr}$, centering would take
no more than a year. However, 
for $\mh\simeq\mz$ a factor of 50 more $\ltot$ is required just for centering
because of the large $Z\to b\anti b$ background.
Thus, for the anticipated $\call$ the \muc\ is not useful if 
the Higgs boson mass is too close to $\mz$.

Once centered, we will wish to measure with precision:
(i) the very tiny Higgs width ---
$\gamh=1-10\mev$ for a SM-like Higgs with $\mh\lsim 140\gev$;
(ii) $\sigma(\mupmum\to \h\to X)$ for 
$X=\tauptaum, b\anti b,c\anti c,W\wstar,Z\zstar$.                  
The accuracy achievable was studied in Ref.~\cite{bbgh}. The three-point
scan of the Higgs resonance described there is the optimal procedure for
performing both measurements simultaneously. We summarize
the resulting statistical
errors in the case of a SM-like $\h$ with $\mh=110\gev$, assuming $R=0.003\%$
and an integrated (4 to 5 year) $\ltot=0.4\fbi$.~\footnote{For $\sigma\br$
measurements, $\ltot$ devoted to the optimized three-point scan 
is equivalent to $\sim\ltot/2$ at the $\rts=\mh$ peak.}
One finds $1\sigma$ errors for $\sigma\br(X)$ of $8,3,22,15,190\%$
for the $X=\tauptaum,b\anti b, c\anti c,W\wstar,Z\zstar$ channels,
respectively, and a $\gamh$ error of 16\%. The individual
channel $X$ results assume
the $\tau,b,c$ tagging efficiencies described in Ref.~\cite{bking}.
We now consider how useful measurements at these accuracy
levels will be.

If only $s$-channel Higgs factory \muc\ data are available
(\ie\ no $Z\h$ data from an \ec\ or \muc), then the 
$\sigma\br$ ratios (equivalently squared-coupling ratios)
that will be most effective for 
discriminating between the SM Higgs
boson and a SM-like Higgs boson such as the $\hl$ of supersymmetry are
${(W\wstar\h)^2\over (b\anti b\h)^2},$
${(c\anti c\h)^2\over (b\anti b\h)^2},$
${(W\wstar\h)^2\over (\tauptaum\h)^2},$ and
${(c\anti c\h)^2\over (\tauptaum\h)^2}.$
The $1\sigma$ errors (assuming $\ltot=0.4\fbi$ 
in the three-point scan centered on $\mh=110\gev$, or $\ltot=0.2\fbi$
with $\rts=\mh=110\gev$)
for these four ratios are $15\%$, $20\%$, $18\%$ and $22\%$, respectively.
Systematic  errors for $(c\anti c\h)^2$ and $(b\anti b\h)^2$
of order $5\%-10\%$
from uncertainty in the $c$ and $b$ quark mass will also enter.
In order to interpret these errors one must compute
the amount by which the above ratios differ in the minimal
supersymmetric model (MSSM) vs. the SM for $\mhl=\mhsm$. The percentage
difference turns out to be essentially identical for
all the above ratios and is a function almost only of the MSSM Higgs sector
parameter $\mha$, with very little dependence on $\tanb$ or top-squark
mixing. At $\mha=250\gev$ ($420\gev$) one finds MSSM/SM $\sim 0.5$ ($\sim
0.8$). Combining the four independent ratio measurements 
and including the systematic errors, one concludes that 
a $>2\sigma$ deviation from the SM predictions would
be found if the observed $110\gev$ Higgs is the MSSM $\hl$ and $\mha < 400\gev$.
Note that the magnitude of the deviation would provide
a determination of $\mha$.

\begin{figure}[h]
\begin{center}
{\epsfig{file=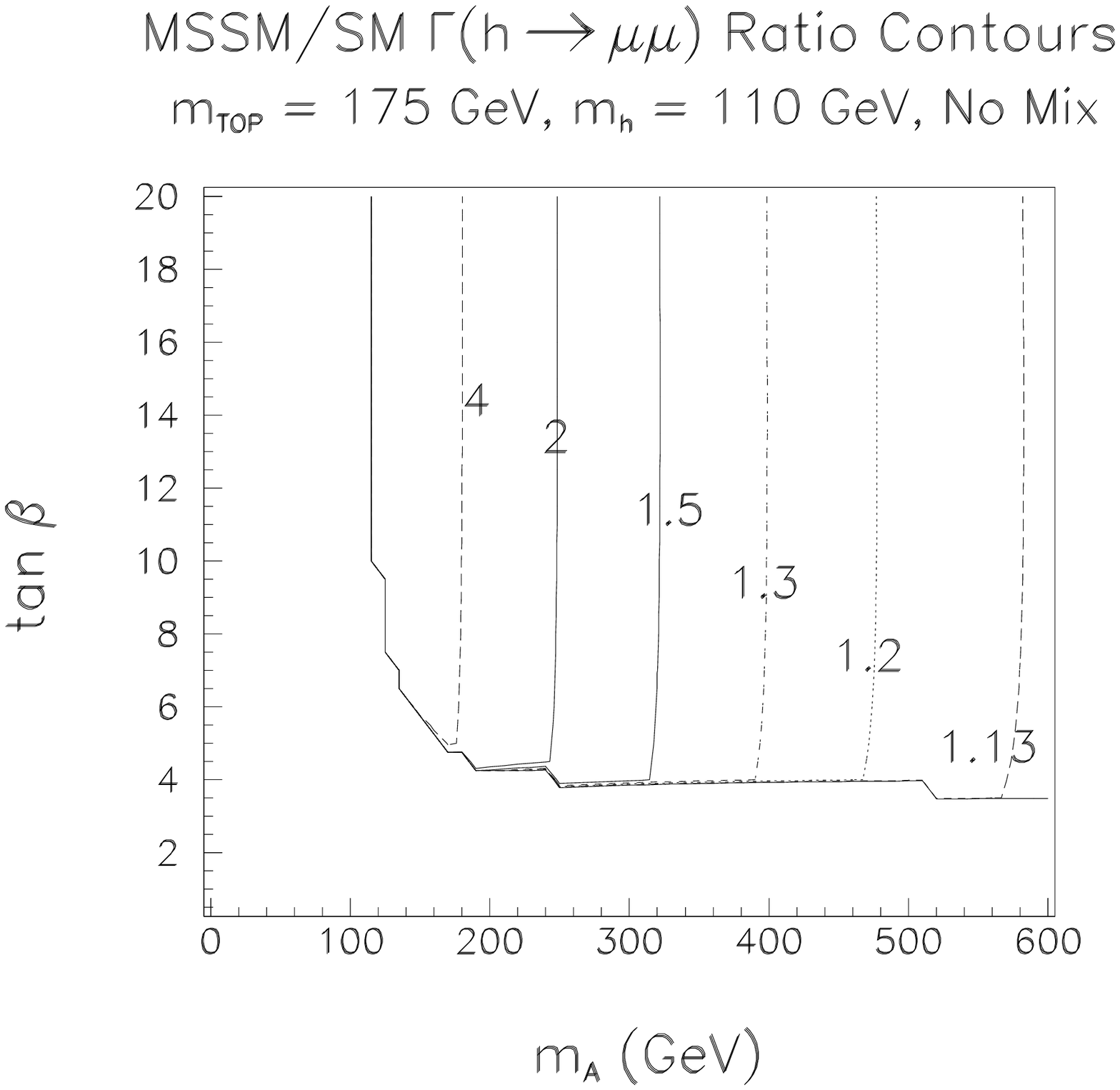,width=3.5in}}
\caption{We give $(\mha,\tanb)$ parameter space contours
for ${\Gamma(\hl\to\mupmum)\over \Gamma(\hsm\to\mupmum)}$:
no-squark-mixing, $\mhl,\mhsm=110\gev$. \label{mumucontours}}
\end{center}
\end{figure}

If, in addition to the $s$-channel measurements we also have \ellc\
$\rts=500\gev$, $\ltot=200\fbi$ data, it will be possible
to discriminate at an even more accurate level between the $\hl$
and the $\hsm$.  The most powerful technique for doing so employs
the four determinations of $\Gamma(\h\to\mupmum)$ below:
\bea
&{[\Gamma(\h\to\mupmum)\br(\h\to
b\anti b)]_{\mu{\rm C}}\over \br(\h\to b\anti b)_{\ell{\rm C}}};
~~~~~~~~~~~~~~~~
{[\Gamma(\h\to\mupmum)\br(\h\to
W\wstar)]_{\mu{\rm C}}\over\br(\h\to W\wstar)_{\ell{\rm C}}};&\nn\\
&{[\Gamma(\h\to\mupmum)\br(\h\to
Z\zstar)]_{\mu{\rm C}}[\gamh]_{\mu{\rm C}+\ell{\rm C}}
\over\Gamma(\h\to Z\zstar)_{\ell{\rm C}}};
~~~
{[\Gamma(\h\to\mupmum)\br(\h\to
W\wstar)\gamh]_{\mu{\rm C}}\over\Gamma(\h\to W\wstar)_{\ell{\rm C}}}.&
\label{gammumu}
\eea
The resulting $1\sigma$ error for $\Gamma(\h\to\mupmum)$ is $\lsim 5\%$.
Fig.~\ref{mumucontours}, which plots the ratio of
the $\hl$ to $\hsm$ partial width in $(\mha,\tanb)$ parameter
space for $\mhl=\mhsm=110\gev$, shows that this level of error allows one
to distinguish between the $\hl$ and $\hsm$ at the $3\sigma$ level 
out to $\mha\gsim 600\gev$. 
Additional advantages of a $\Gamma(\h\to\mupmum)$ measurement are:
(i) there are no systematic uncertainties arising from uncertainty
in the muon mass; (ii) the error on
$\Gamma(\h\to\mupmum)$ increases only very slowly as the $s$-channel
$\ltot$ decreases,~\footnote{This is because
the $\Gamma(\h\to\mupmum)$ error is dominated by the $\rts=500\gev$
measurement errors.} in contrast to the errors for the previously
discussed ratios of branching ratios from the \muc\
$s$-channel data which scale as $1/\sqrt{\ltot}$.
Finally, we note that $\gamh$ alone cannot be used
to distinguish between the MSSM $\hl$ and SM $\hsm$ in a model-independent way. 
Not only is the error substantial ($\sim 12\%$ if we combine \muc, $L=0.4\fbi$ 
$s$-channel data with \ellc, $L=200\fbi$ data)
but also $\gamh$ depends on many things, including (in the MSSM) the
squark-mixing model. Still, deviations from SM predictions are generally
substantial if $\mha\lsim 500\gev$ implying that the measurement of $\gamh$
could be very revealing.

We note that
the above errors and results hold approximately for all $\mh \lsim 150\gev$
so long as $\mh$ is not too close to $\mz$.

Precise measurements of the couplings of the SM-like Higgs boson could
reveal many other types of new physics. For example, if a significant
fraction of a fermion's mass is generated radiatively (as opposed
to arising at tree-level), then the $\h f\anti f$ coupling 
and associated partial width will deviate from SM expectations
\cite{borzumati}. Deviations of order 5\% to 10\% (or more)
in $\Gamma(\h\to \mupmum)$
are quite possible and, as discussed above, potentially detectable.

\medskip
\noindent{\bf\boldmath The MSSM $\hh$, $\ha$ and $\hpm$}
\medskip

We begin by recalling \cite{jfghiggs} that
the possibilities for $\hh,\ha$ discovery are limited at other machines.
(i) Discovery of $\hh,\ha$ is not possible at the LHC
for all $(\mha,\tanb)$: \eg\ if $\mstop=1\tev$, consistency
with the observed value of $\br(b\to s\gam)$
requires $\mha>350\gev$, in which case the LHC might
not be able to detect the $\hh,\ha$ at all, and certainly not
for all $\tanb$ values. 
If $\tanb\lsim 3$, detection might be possible in the $\hh,\ha\to t\anti t$
final state, but would require $\lsim 10\%$ systematic
uncertainty in understanding the absolute normalization
of the $t\anti t$ background.  Otherwise, and certainly for $\tanb\gsim 3$, 
one must employ $b\anti b\ha,b\anti b\hh$ associated 
production, first analyzed in Refs.~\cite{dgv,dgvnew} and recently
explored further in \cite{froidrw,cpyuan}. There is currently considerable
debate as to what portion of $(\mha,\tanb)$ parameter space can be covered
using the associated production modes.
In the update of \cite{dgvnew}, it is claimed that 
$\tanb\gsim 5$ ($\gsim 15$) is required for $\mha\sim 200\gev$ ($\sim
500\gev$). Ref.~\cite{froidrw} claims that still higher
$\tanb$ values are required, $\tanb\gsim 20$ ($\tanb\gsim 30$),
whereas Ref.~\cite{cpyuan} claims
$\tanb\gsim 2 $ ($\gsim 4$) will be adequate.
(ii) At $\rts=500\gev$, $\epem\to \hh\ha$ 
pair production probes only to $\mha\sim\mhh\lsim 230-240\gev$.
(iii)
A $\gam\gam$ collider could potentially probe up
to $\mha\sim\mhh\sim 0.8 \rts\sim 400\gev$, but only
for $\ltot\gsim 150-200\fbi$ \cite{ghgamgam}.

Thus, it is noteworthy that 
$\mupmum\to\hh,\ha$ in the $s$-channel 
potentially allows production and study
of the $\hh,\ha$ up to $\mha\sim\mhh\lsim \rts$. 
To assess the potential, let us (optimistically) 
assume that a total of $\ltot=50\fbi$ (5 yrs
running at $<\call>=1\times 10^{33}$) can be accumulated for 
$\rts$ in the $ 250-500\gev$ range. (We note that $\gamha$ and $\gamhh$,
although not big, are of a size such that resolution of $R\gsim 0.1\%$
will be adequate to maximize the $s$-channel cross section, thus
allowing for substantial $\call$.)

\begin{figure}[h]
\begin{center}
{\epsfig{file=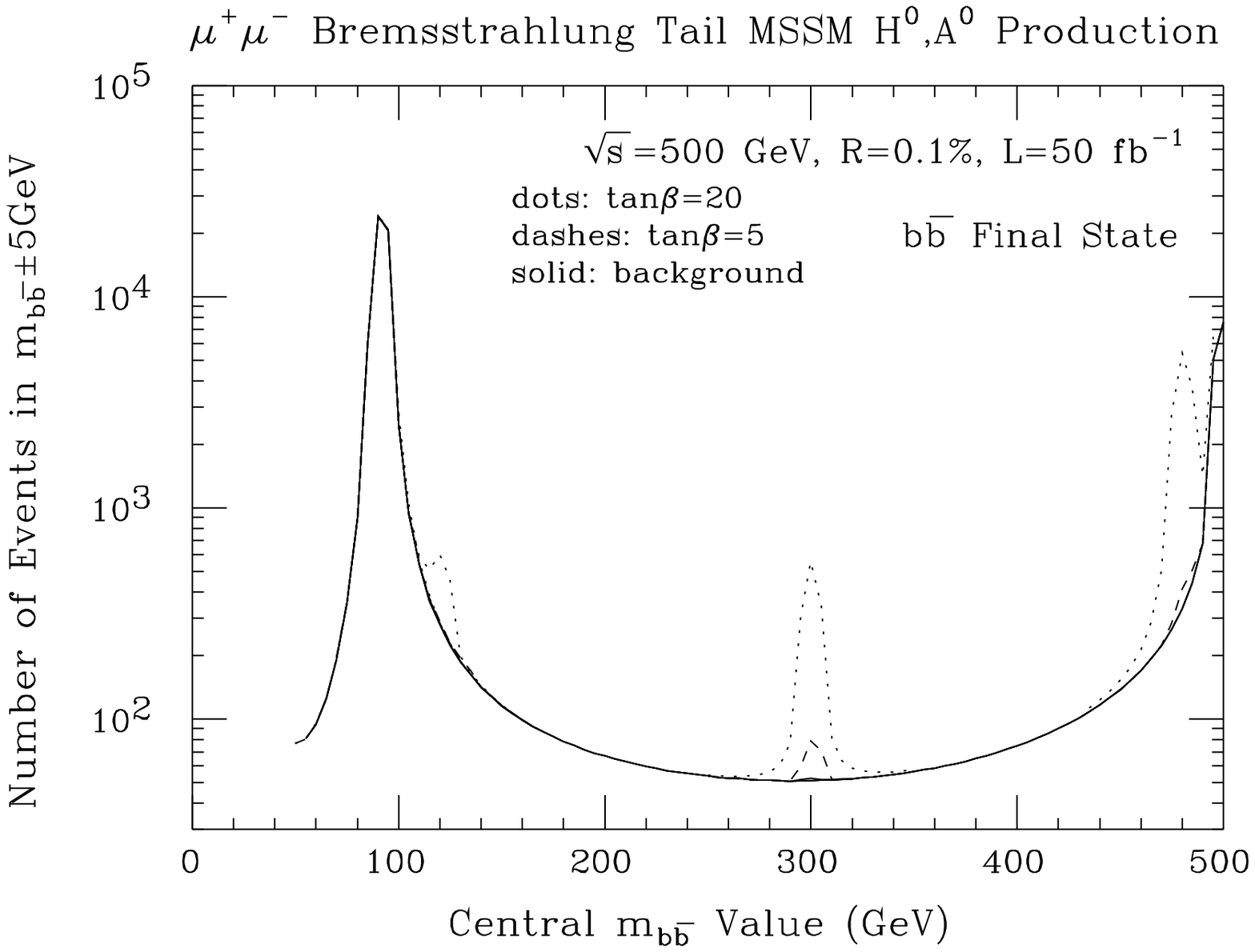,width=3.5in}} 
\caption{ $N(b\anti b)$ in the $\mbb\pm5\gev$ interval
vs. $\mbb$ for $\protect\rts=500\gev$, $\ltot=50\fbi$,
and $R=0.1\%$: peaks are shown for
$\mha=120$, $300$ or $480\gev$, with $\tanb=5$ and $20$ in each case.
 \label{mbbscan}}
\end{center}
\end{figure}

There are then several possible scenarios.
(a) If we have some preknowledge or restrictions
on $\mha$ from LHC discovery or 
from $s$-channel measurements of $\hl$ properties,
then $\mupmum\to\hh$ and $\mupmum\to\ha$ can be studied with
precision for all $\tanb\gsim 1-2$.
(b) 
If we have no knowledge of $\mha$ other than $\mha\gsim 250-300\gev$
from LHC, then we might wish to search for the $\ha,\hh$
in $\mupmum\to\hh,\ha$ by scanning over $\rts=250-500\gev$.
If their masses lie in this mass range, then their discovery
by scanning will be possible for most of $(\mha,\tanb)$ parameter
space such that they cannot be discovered at the LHC (in particular,
if $\mha\gsim 250\gev$ and $\tanb\gsim 4-5$).
(c) Alternatively, if the \muc\
is simply run at $\rts=500\gev$ and $\ltot\sim 50\fbi$ is accumulated,
then $\hh,\ha$ in the $250-500\gev$ mass range
can be discovered in the $\rts$ bremsstrahlung tail if the $b\anti b$
mass resolution (either by direct reconstruction or
hard photon recoil) is of order $\pm 5\gev$ and if $\tanb\gsim 6-7$ (depending
on $\mha$). Typical peaks are illustrated in 
Fig.~\ref{mbbscan}.~\footnote{SUSY decays are assumed to be absent in
this and the following figure.}

\begin{figure}[h]
\begin{center}
{\epsfig{file=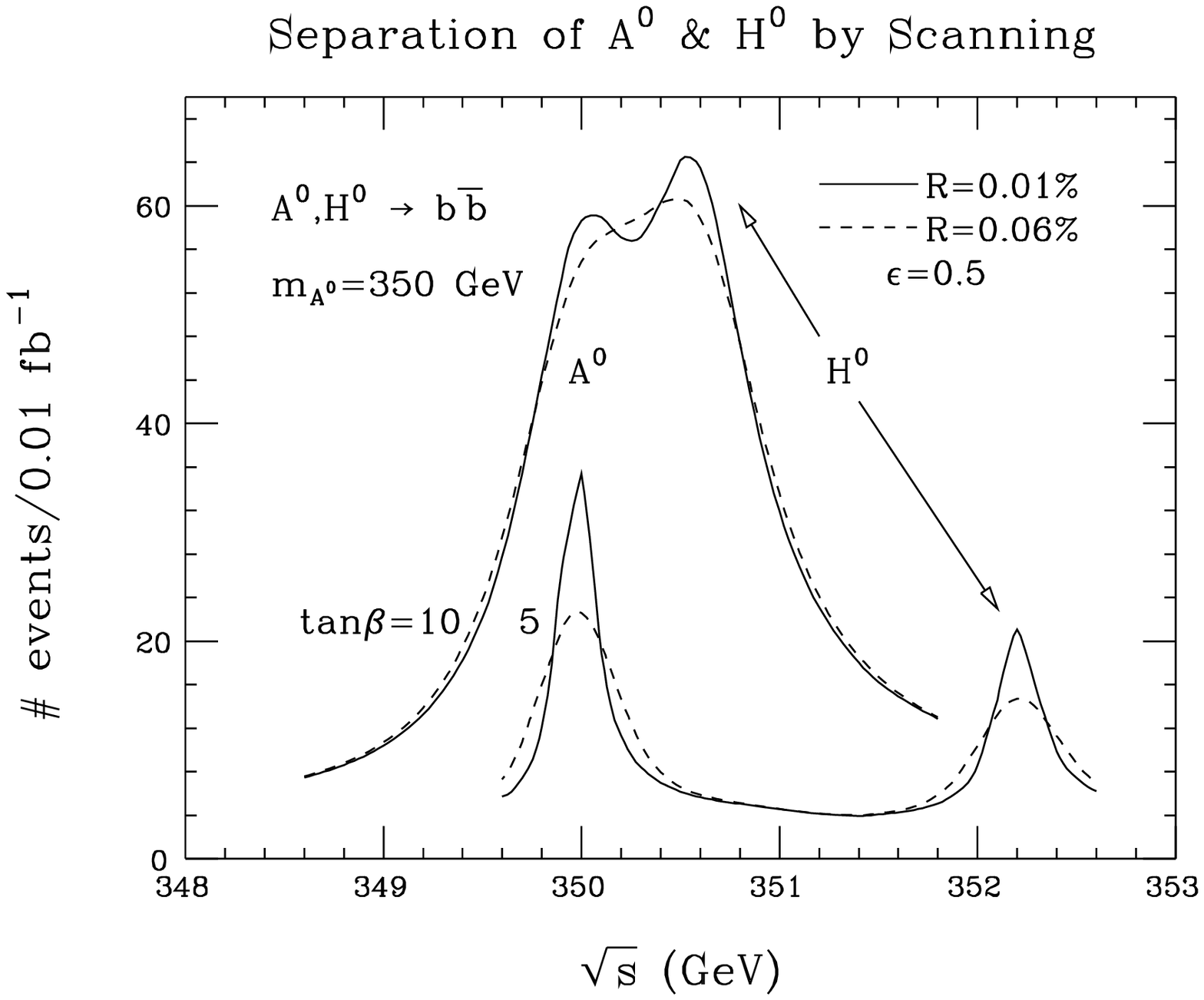,width=3.5in}}
\caption{ $N(b\anti b)$ (for $0.01\fbi$)
vs. $\protect\rts$, for $\mha=350\gev$ $\hh,\ha$ resonance (with
$\tanb=5$ and 10), including the $b\anti b$ continuum background.
\label{hhhasusyrtsscan}}
\end{center}
\end{figure}

Finally, once the closely degenerate $\ha,\hh$ are discovered,
it will be extremely interesting to be able to separate the resonance
peaks.  This will probably only be possible at a muon collider
with small $R\lsim 0.01\%$ if $\tanb$ is large, as illustrated in
Fig.~\ref{hhhasusyrtsscan}.

We note that the above results assume that SUSY decays of the $\hh$ and $\ha$
do not have a large net branching ratio for $\mha,\mhh\lsim 500\gev$. If SUSY
decays are significant, the possibilities and strategies for $\hh,\ha$ discovery
at all machines would have to be re-evaluated.

We end this sub-section
with just a few remarks on the possibilities for production
of $\hh\ha$ and $\hp\hm$ pairs at a high energy \muc\ (or \ec).
Since $\mha\gsim 1\tev$ cannot be ruled out simply on the basis
of hierarchy and naturalness (although fine-tuning is stretched),
it is possible that energies of $\rts >2\tev$ could be
required for pair production.
If available, then it has been shown \cite{gk,fengmoroi} that
discovery of $\hh\ha$ in their $b\anti b$ or $t\anti t$ 
decay modes and $\hp\hm$ in their $t\anti b$ and $b\anti t$ decays
will be easy for expected luminosities, even if SUSY decays are
present. As a by-product, the masses will be measured
with reasonable accuracy. 

Regardless of whether we see the $\hh,\ha$ in $s$-channel production
or via pair production, one can measure
branching ratios to other channels, including supersymmetric pair
decay channels with good accuracy.
In fact, the ratios of branching ratios and the value
of $\mha\sim\mhh\sim\mhpm$ will be measured with sufficient
accuracy that, in combination with one gaugino mass, say the chargino
mass (which will also presumably be well-measured)
it will be possible \cite{gk} to discriminate 
with incredible statistical significance between
different closely similar GUT scenarios for the 
GUT-scale soft-supersymmetry-breaking masses.
Thus, Higgs pair production could be very valuable in the
ultimate goal of determining all the soft-SUSY-breaking parameters.

Finally, entirely unexpected decays of the heavy Higgs bosons of SUSY
(or other extended Higgs sector) could be present. For example,
non-negligible branching ratios for $\hh,\ha\to t\anti c+c\anti t$ FCNC 
decays are not inconsistent with current theoretical model-building ideas
and existing constraints \cite{reina}. The muon collider $s$-channel 
$\mupmum\to \hh,\ha$ event rate is sufficient to probe rather small values
for such FCNC branching ratios.

\medskip
\noindent{\bf Verifying Higgs CP Properties}
\medskip

Once a neutral Higgs boson is discovered, determination of its CP nature
will be of great interest. For example,
direct verification that the SM Higgs is CP-even would 
be highly desirable.  Indeed, if a neutral Higgs boson is found to
have a mixed CP nature (implying CP violation in the Higgs sector),
then neither the SM nor the MSSM can be correct. In the case of the SM,
one must have a multi-doublet (or more complicated) Higgs sector.
In the case of the MSSM, at least a singlet Higgs boson (as in the NMSSM)
would be required to be present in addition to the standard two doublets.

One finds that the $\gam\gam$ and $\mupmum$ single Higgs production modes
provide the most elegant and reliable techniques for CP determination.
In $\gam\gam$ collisions at the \ec\ (a $\gam\gam$ collider is not possible
at the \muc), one establishes definite polarizations $\vec e_{1,2}$ for 
the two colliding photons in the photon-photon center of mass. Since
$\call_{\gam\gam\h}=\vec e_1\cdot \vec e_2 \cale+(\vec e_1\times \vec e_2)_z
\calo$, where $\cale$ and $\calo$ are of similar size if
the CP-even and CP-odd (respectively) components of the $\h$ are comparable.
There are two important types of measurement.
The first \cite{ggcpgamgam}
is the difference in rates for photons colliding with $++$ vs. $--$ 
helicities, which is non-zero only if CP violation is present.
Experimentally, this difference can be measured by 
simultaneously flipping the helicities of both of the initiating
back-scattered laser beams.
The second \cite{ggcpgamgam,gkgamgamcp,kksz} is 
the dependence of the $\h$ production rate on the relative orientation of
transverse polarizations $\vec e_1$ and $\vec e_2$ for the two colliding
photons.
In the case of a CP-conserving Higgs sector, the production rate is 
maximum for a CP-even (CP-odd) Higgs boson when $\vec e_1$ is parallel
(perpendicular) to $\vec e_2$. The limited transverse polarization
that can be achieved at a $\gam\gam$ collider implies that very high
luminosity is needed for such a study.

In the end, a $\mupmum$ collider might well prove to be the best machine
for directly probing the CP properties of a Higgs boson that
can be produced and detected in the $s$-channel mode \cite{atsoncp,dpfreport}.
Consider transversely polarized muon beams.
For 100\% transverse polarization and an angle $\phi$ between
the $\mu^+$ transverse polarization and the $\mu^-$ transverse polarization,
one finds
\begin{equation}
\sigma(\phi)\propto 1 - {a^2-b^2\over a^2+b^2} \cos\phi 
+ {2ab\over a^2+b^2}\sin\phi\,,
\label{cpmu}
\end{equation}
where the coupling of the $\h$ to muons is given by 
$\h\anti \mu(a+ib\gamma_5)\mu$,
$a$ and $b$ being the CP-even and CP-odd couplings, respectively.
If the $\h$ is a CP mixture, both $a$ and $b$ are non-zero and the asymmetry
\begin{equation}
A_1\equiv {\sigma(\pi/2)-\sigma(-\pi/2)\over \sigma(\pi/2)+\sigma(-\pi/2)}
= {2ab\over a^2+b^2}
\end{equation}
will be large. For a pure CP eigenstate the cross section difference
\begin{equation}
A_2\equiv {\sigma(\pi)-\sigma(0) \over \sigma(\pi)+\sigma(0)}
= {a^2-b^2\over a^2+b^2}
\end{equation}
is $+1$ or $-1$ for a CP-even or CP-odd $\h$, respectively.
Since background processes and partial transverse polarization
will dilute the statistics, further study will be needed to
fully assess the statistical level of CP determination
that can be achieved in various cases.

\medskip
\noindent{\bf Exotic Higgs Bosons}
\medskip

If there are doubly-charged Higgs bosons, $\em\em\to \dmm$
probes $\lam_{ee}$ and $\mu^-\mu^-\to \dmm$ probes $\lam_{\mu\mu}$,
where the $\lam$'s are the strengths of the Majorana-like couplings
\cite{jfgemem,frampton,cuypers}.  
Current $\lam_{ee,\mu\mu}$ 
limits are such that factory-like production of a $\dmm$ 
is possible if $\gamdmm$ is small.
Further, a $\dmm$ with $\mdmm\lsim 500-1000\gev$
will be seen previously at the LHC 
(for $\mdmm\lsim 200-250\gev$ at TeV33) \cite{kpitts}.
For small $\lam_{ee,\mu\mu,\tau\tau}$ in the range that
would be appropriate, for example, for the $\dmm$ 
in the left-right symmetric model see-saw
neutrino mass generation context, it may be that
$\gamdmm\ll\sigrts$,~\footnote{For small $\lam_{ee,\mu\mu,\tau\tau}$,
$\gamdmm$ is very small
if the $\dmm\to\wm\wm$ coupling strength is very small or zero, as required
to avoid naturalness problems for $\rho=\mw^2/[\cos^2\theta_w\mz]^2$.}
leading to
$\overline\sigma_{\ell^-\ell^-\to\dmm}\propto \lam_{\ell\ell}^2/\sigrts$.
Note that the absolute rate for $\ell^-\ell^-\to\dmm$
yields a direct determination
of $\lam_{\ell\ell}^2$, which, for a $\dmm$ with very small $\gamdmm$, will
be impossible to determine by any other means. The relative branching
ratios for $\dmm\to e^-e^-,\mu^-\mu^-,\tau^-\tau^-$ will then yield values
for the remaining $\lam_{\ell\ell}^2$'s.
Because of the very small $R=0.003\%-0.01\%$ achievable at a muon collider,
$\mu^-\mu^-$ collisions will probe weaker $\lam_{\mu\mu}$ coupling
than the $\lam_{ee}$ coupling that can be probed in $e^-e^-$ collisions.
In addition, it is natural to anticipate that $\lam_{\mu\mu}^2\gg\lam_{ee}^2$.
A more complete review of this topic is given in Ref.~\cite{jfgememnew}.

\section*{\bf\boldmath Probes of narrow technicolor resonances}

In this section, I briefly
summarize the ability of a low-energy muon collider to observe
the pseudo-Nambu-Goldstone bosons (PNGB's) of an extended technicolor
theory.  These narrow states
need not have appeared at an observable
level in $Z$ decays at LEP. Some of the PNGB's have
substantial $\mupmum$ couplings. Thus, a muon collider search for
them will bear a close resemblance to the light Higgs
case discussed already.  The main difference
is that, assuming they have not been detected ahead of time, we must search
over the full expected mass range.

\begin{figure}[h] 
\begin{center}
\epsfig{file=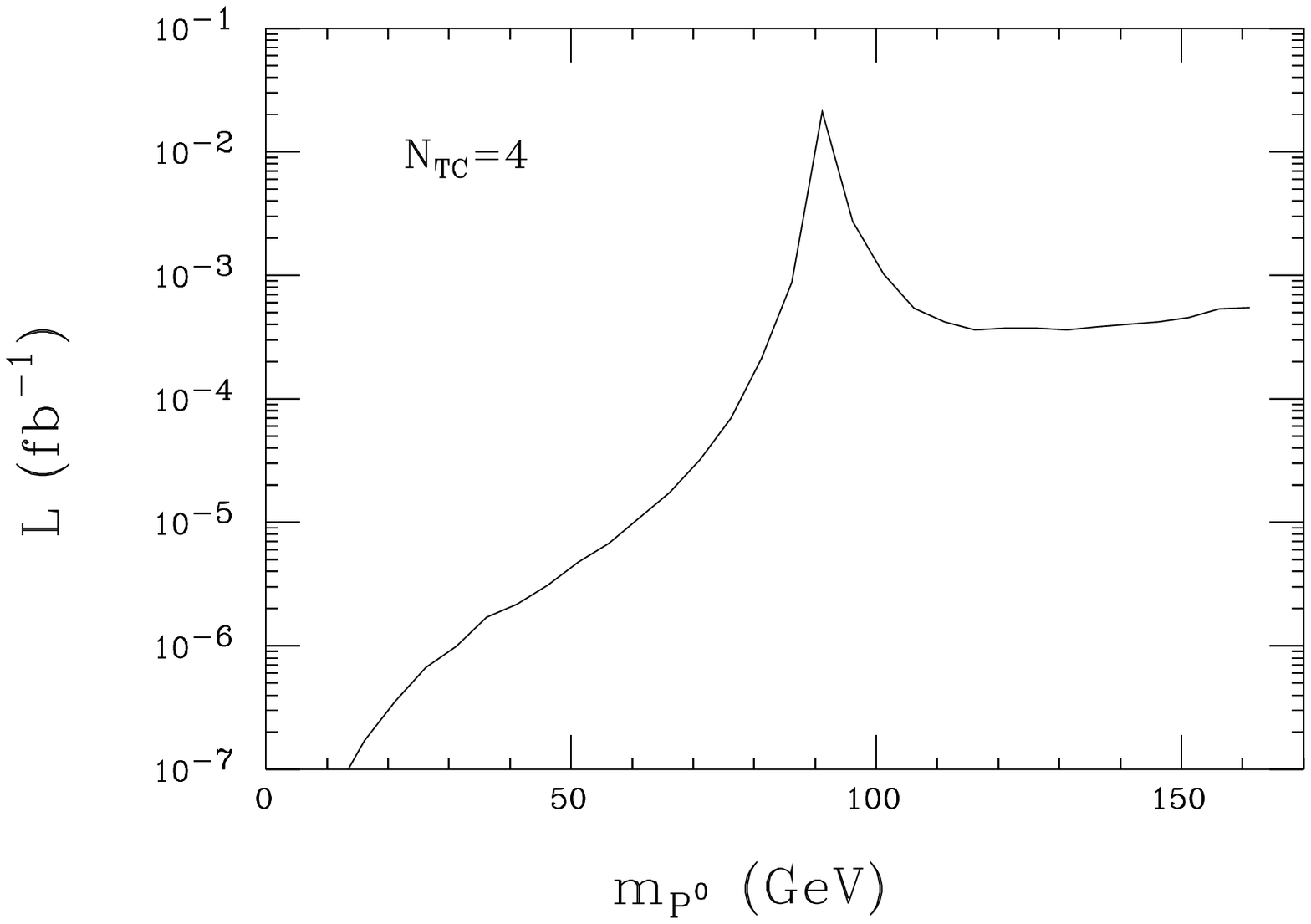,width=3.5in}
\caption{$\ltot$ required 
for a 5$\sigma$ $\pzero$ signal at $\protect\rts=\mpzero$.
\label{figlum}}
\end{center}
\end{figure}

The first results for PNGB's at a muon collider appear
in Refs.~\cite{dominici} and \cite{lane}.
Here I summarize the results for the lightest $\pzero$ PNGB
as given in Ref.~\cite{dominici}. 
Although the specific $\pzero$ properties employed are those predicted
by the extended BESS model \cite{dominici}, they will be
representative of what would be found in any extended technicolor model for
a strongly interacting electroweak sector. The first point
is that $\mpzero$ is expected to be small; $\mpzero\lsim 80\gev$ 
is preferred in the BESS model. 
Second, the Yukawa couplings and
branching ratios of the $\pzero$ are easily determined.
In the BESS model,
${\cal L}_Y= -i \sum_f\lambda_f \bar f\gamma_5 f \pzero$
with
$\lambda_b=\sqrt{\frac 2 3 }\frac {m_b} v$,
$\lambda_\tau=-\sqrt{6}\frac {m_\tau} v$,
$\lambda_\mu=-\sqrt{6} \frac {m_\mu} v$.
Note the sizeable $\mupmum$ coupling.
The $\pzero$ couplings to $\gamma\gamma$ and $gg$
from the ABJ anomaly are also important.
Overall, these couplings are not unlike
those of a light Higgs boson.  Not surprisingly, therefore,
$\gampzero$ is very tiny: $\gampzero=0.2,4,10\mev$ for $\mpzero=10,80,150\gev$,
respectively, for $N_{TC}=4$ technicolor flavors. For such narrow
widths, it will be best to use $R=0.003\%$ beam energy resolution.

\begin{figure}[h]
\begin{center}
\epsfig{file=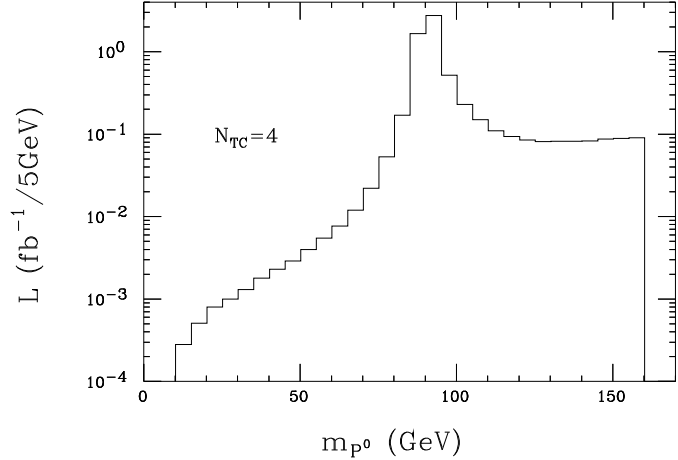,width=3.5in}
\caption{$\ltot$ required to scan indicated 5 GeV intervals
and either discover or eliminate the $\pzero$ at the $3\sigma$
level. \label{lumtable}}
\end{center}
\end{figure}
For the detailed tagging efficiencies \etc\ 
described in \cite{dominici}, the $\ltot$
required to achieve $\sum_k S_k/\sqrt{\sum_k B_k}=5$
at $\rts=\mpzero$, after summing over the optimal selection of
the $k=b\bar b$, $\tau^+\tau^-$, $c\bar c$, and $gg$ channels
(as defined after tagging), is plotted in Fig.~\ref{figlum}. 
Very modest $\ltot$ is needed unless $\mpzero\sim\mz$.
Of course, if we do not have any information regarding the $\pzero$ mass,
we must scan for the resonance. The (very conservative, see \cite{dominici}
for details) estimate for the luminosity required for scanning a 
given 5 GeV interval and either discovering
or eliminating the $\pzero$ in that interval
at the $3\sigma$ level is plotted in Fig.~\ref{lumtable}.
If the $\pzero$ is as light as expected in the extended BESS model, then
the prospects for discovery by scanning would be excellent. For example,
a $\pzero$ lying in the $\sim 10\gev$ to $\sim 75\gev$ mass interval
can be either discovered or eliminated at the $3\sigma$ level with
just $0.11\fbi$ of total luminosity, distributed in proportion to
the luminosities plotted in Fig.~\ref{lumtable}. 
The $\call$ that
could be achieved at these low masses is being studied \cite{palmer}.
A $\pzero$
with $\mpzero\sim\mz$ would be much more difficult to discover
unless its mass was approximately known. A $3\sigma$ scan
of the mass interval from $\sim 105\gev$ to $160\gev$ would require
about $1\fbi$ of integrated luminosity, which is more than
could be comfortably achieved for the conservative $R=0.003\%$ $\call$
values assumed at the Fermilab-97 workshop.

\section*{Discussion and Conclusions}

There is little doubt that a variety of accelerators will be needed 
to explore all aspects of the physics that lies beyond the Standard Model
and accumulate adequate luminosity for this purpose in a timely fashion.
For any conceivable new-physics scenario, a muon collider would be a very
valuable machine, both for discovery and detailed studies.
Here we have reviewed the tremendous value of a muon collider for studying
any narrow resonance with $\mupmum$ (or $\mu^-\mu^-$) couplings, 
focusing on neutral light Higgs bosons
and the Higgs-like pseudo-Nambu-Goldstone bosons that would be present
in almost any technicolor model. A muon collider could well provide
the highest statistics determinations of many important Higgs or PNGB 
fundamental couplings. In particular, it might provide the only direct
measurement of the very important $\mupmum$ coupling.
Measurement of this coupling will very possibly
allow discrimination between a SM Higgs boson
and its light $\hl$ SUSY counterpart. Comparison of the $\mupmum$ coupling
to the $\tauptaum$ coupling (one may be able to approximately determine
the latter from branching ratios) will also be of extreme interest.
For Higgs physics, developing machine designs that yield the highest
possible luminosity at low energies, while maintaining excellent
beam energy resolution, should be a priority.

\end{document}